\documentclass[12pt,aps,showpacs]{revtex4}
\usepackage{epsfig}

\newcommand{\nc}{\newcommand}
\nc{\be}{\begin{equation}}
\nc{\ee}{\end{equation}}
\nc{\bea}{\begin{eqnarray}}
\nc{\eea}{\end{eqnarray}}
\nc{\lsim}{\mbox{\raisebox{-.6ex}{~$\stackrel{<}{\sim}$~}}}
\nc{\gsim}{\mbox{\raisebox{-.6ex}{~$\stackrel{>}{\sim}$~}}}
\nc{\gtwid}{\mathrel{\raise.3ex\hbox{$>$\kern-.75em\lower1ex\hbox{$\sim$}}}}
\nc{\ltwid}{\mathrel{\raise.3ex\hbox{$<$\kern-.75em\lower1ex\hbox{$\sim$}}}}
\nc{\comp}{{\rm C}\llap{\vrule height7.1pt width1pt depth-.4pt\phantom t}}

\begin{document}

\vskip -0.2in

\rightline{CERN-TH/2003-173, HD-THEP-03-39, UFIFT-HEP-03-21}

\vskip 0.2in

\title{Production of Massless Fermions during Inflation~\footnote{At the end of the paper
an addendum is appended which contains Erratum for this paper.}}

\author{T. Prokopec}
\email{T.Prokopec@phys.uu.nl}
\affiliation{Institute for Theoretical Physics (ITP) \& Spinoza Institute, Utrecht University,
  Leuvenlaan 4, 3584 TC Utrecht, The Netherlands}

\author{R. P. Woodard}
\email{woodard@phys.ufl.edu}
\affiliation{Department of Physics, University of Florida, Gainesville, 
Florida 32611, USA}


\begin{abstract}
We compute the one loop self energy, in a locally de Sitter background,
for a massless fermion which is Yukawa-coupled to a massless, minimally
coupled scalar. We then solve the modified Dirac equation resulting from
inclusion of the self energy. We find faster-than-exponential growth
in the fermion wave function, consistent with the production of fermions
through a process in which a scalar and a fermion-anti-fermion pair are 
ripped out of the vacuum by inflation.
\end{abstract}


\maketitle

\section{Introduction}

Consider quantum field theory in the presence of a homogeneous, isotropic and
spatially flat geometry,
\begin{equation}
ds^2 = -dt^2 + a^2(t) d\vec{x} \cdot d\vec{x} \; . \label{comoving}
\end{equation}
By virtue of spatial translation invariance we can label particle states
by their constant wave vectors, $\vec{k}$. However, the physical 3-momentum 
associated with this label is $\vec{k}/a(t)$. And if such a particle has 
mass $m$, its energy is,
\begin{equation}
E(t,\vec{k}) = \sqrt{m^2 + \Vert \vec{k} \Vert^2/a^2(t)} \; .
\end{equation}
Now consider the appearance, at time $t$, of a pair of virtual particles with
wave vectors $\pm \vec{k}$. This process conserves 3-momentum but not energy.
According to the energy-time uncertainty principle this violation of energy
conservation will not be detectable provided the pair disappears back into 
the vacuum before a time interval ${\Delta t}$ such that,
\begin{equation}
\int_t^{t + {\Delta t}} \!\!\!\! dt' \, 2 E(t',\vec{k}) \ltwid 1 \; . 
\label{unc}
\end{equation}
For any $m \neq 0$ the integral grows without bound, so ${\Delta t}$ is finite 
for massive particles. However, if $m=0$ and the scale factor grows rapidly 
enough, the integral remains bounded even for ${\Delta t} \longrightarrow
\infty$. This means that massless virtual particles need never recombine.

``Rapidly enough'' turns out to mean $\ddot{a}(t) > 0$, which is a type of
expansion known as ``inflation'' \cite{Linde}. Whether or not inflation leads 
to a significant amount of particle production for a given species depends 
upon the rate at which virtual particles of that species emerge from the 
vacuum. It turns out that the emergence rate for particles which are 
conformally invariant falls like $1/a(t)$ \cite{PW1}. This means that any 
conformally invariant and massless virtual particles which happen to emerge 
from the vacuum will become real, but not many emerge. Hence there is 
negligible direct production of conformally invariant particles during 
inflation.

Most familiar massless particles possess conformal invariance. In particular,
photons are conformally invariant in $D=4$ spacetime dimensions, and massless
Dirac fermions are conformally invariant in any dimension. On the other hand,
neither massless, minimally coupled scalars nor gravitons are conformally
invariant, and both are produced copiously during inflation. This effect is
observable; the production of nearly massless, minimally coupled scalars is 
believed to be responsible for the primordial anisotropies in the cosmic ray 
microwave background \cite{Slava,WMAP}.

One naturally wonders what happens when a massless but conformally invariant 
particle interacts with a massless particle which is not conformally invariant. 
The first step in answering this question is to compute the one particle 
irreducible (1PI) 2-point function of the conformally invariant field. One 
next uses this 1PI 2-point function to correct the linearized equations of 
motion. The effect of inflation on the conformally invariant particle can be 
inferred by solving these equations for the plane wave mode functions.

We have previously carried out such a study for photons interacting with 
a charged, massless and minimally coupled scalar \cite{PTW1,PTW2}. The result 
is that the mode functions of super-horizon wave vectors oscillate around a
nonzero constant with increasing frequency and decreasing amplitude 
\cite{PW2}. This implies zero photon production and an enormous enhancement 
in the 0-point energy, roughly similar to what happens for massive photons
in Proca electrodynamics. The physical interpretation is that the inflationary 
production of charged scalars results in a diffuse plasma which inhibits the 
propagation of light.

The purpose of this paper is to study what happens when Dirac fermions are
made to Yukawa-interact with a real, massless and minimally coupled scalar. 
We will show that although the fermion mass remains zero, the mode functions 
for super-horizon wave vectors grow in a manner that is consistent with the 
production of fermions. The physical interpretation seems to be that inflation 
alters the constraint of energy conservation to permit the spontaneous 
appearance of a scalar and a fermion-anti-fermion pair.

In Section II we compute the one loop fermion self energy. In Section III this
is used to correct the Dirac equation. Although this equation is nonlocal,
an asymptotic form is derived for the late time behavior of a spatial plane 
wave. In Section IV we show that the particle production inferred from the 
mode functions is consistent with a 3-particle creation process. Our results 
are summarized and discussed in Section V.

\section{The one loop self energy}

We begin by reviewing the conventions appropriate to Dirac fields in a
nontrivial geometry. In order to facilitate dimensional regularization 
we make no assumption about the spacetime dimension $D$. The gamma matrices 
$\gamma^{b}_{ij}$ ($b=0,1,\dots, D\!-\!1$) anti-commute in the usual way, 
$\{\gamma^b,\gamma^c\} = -2 \eta^{bc} I$. One interpolates between local 
Lorentz indices ($b,c,d,\dots$) and vector indices (lower case Greek letters) 
with the vierbein field, $e_{\mu b}(x)$. The metric is obtained by contracting 
two vierbeins with the Minkowski metric, $g_{\mu\nu}(x) = e_{\mu b}(x) 
e_{\nu c}(x) \eta^{bc}$. The vierbein's vector index is raised and lowered by 
the metric ($e^{\mu}_{~b} = g^{\mu\nu} e_{\nu b}$) while the local Lorentz
index is raised and lowered with the Minkowski metric ($e_{\mu}^{~b} = 
\eta^{bc} e_{\mu c}$). The spin connection and the Lorentz representation 
matrices are,
\begin{equation}
A_{\mu bc} \equiv e^{\nu}_{~b} \Bigl(e_{\nu c , \mu} - \Gamma^{\rho}_{~\mu \nu}
e_{\rho c}\Bigr) \qquad , \qquad J^{bc} \equiv \frac{i}4 [\gamma^b,\gamma^c]
\; .
\end{equation}

Let $\phi(x)$ represent a real scalar field and let $\psi_i(x)$ stand for a
Dirac field. In a general background metric the Lagrangian we wish to study 
would be,
\begin{equation}
{\cal L}_{\rm gen} = -\frac12 \partial_{\mu} \phi \partial_{\nu} \phi 
g^{\mu\nu} \sqrt{-g} + \overline{\psi} e^{\mu}_{~b} \gamma^b \Bigl(i 
\partial_{\mu} - \frac12 A_{\mu cd} J^{cd} \Bigr) \psi \sqrt{-g} - f \phi 
\overline{\psi} \psi \sqrt{-g} \; ,
\end{equation}
where $\overline{\psi} \equiv \psi^{\dagger} \gamma^0$ is the usual Dirac
adjoint and $f$ is the Yukawa coupling constant. The geometry of interest 
is the very special form associated with the homogeneous and isotropic 
element (\ref{comoving}). By defining a new time coordinate $d\eta \equiv 
dt/a(t)$, the metric of this geometry can be made conformal to the Minkowski 
metric,
\begin{equation}
ds^2 = a^2 \Bigl(-d\eta^2 + d\vec{x} \cdot d\vec{x} \Bigr) \; . \label{conf}
\end{equation}
A convenient choice for the associated vierbein is, $e_{\mu b} = a 
\eta_{\mu b}$. With these simplifications the spin connection assumes the 
form,
\begin{equation}
e_{\mu b} = a \eta_{\mu b} \Longrightarrow A_{\mu cd} = \Bigl(\eta_{\mu c} 
\partial_d - \eta_{\mu d} \partial_c \Bigr) \ln(a) \; .
\end{equation}
And our Lagrangian reduces to,
\begin{equation}
{\cal L}_{\rm conf} = -\frac12 a^{D\!-\!2} \partial_{\mu} \phi \partial_{\nu} 
\phi \eta^{\mu\nu} + \Bigl(a^{\frac{D-1}2} \overline{\psi}\Bigr) i \gamma^{\mu}
\partial_{\mu} \Bigl(a^{\frac{D-1}2} \psi \Bigr) - f a^D \phi 
\overline{\psi} \psi \; . \label{L}
\end{equation}

Having defined the Lagrangian we now give the propagators and vertices which 
comprise the position space Feynman rules. Of course vertices are always 
straightforward. The $\phi \overline{\psi}_i \psi_j$ vertex and the fermion 
field strength renormalization are,
\begin{equation}
-i f a^D \delta_{ij} \qquad {\rm and} \qquad i {\delta Z}_2 
(a a')^{\frac{D-1}2} \gamma^{\mu}_{ij} \, i \partial_{\mu} \delta^D(x-x') \; ,
\label{verts}
\end{equation}
where $a \equiv a(t)$ and $a' \equiv a(t')$ are the scale factors evaluated
at $x^0 = \eta$ and $x^{\prime 0} = \eta'$, respectively. We do not need 
scalar mass and field strength renormalization.

The fermion propagator is best expressed in terms of the conformal coordinate 
interval,
\begin{equation}
{\Delta x}^2(x;x') \equiv \Vert \vec{x} \!-\! \vec{x}' \Vert^2 - (\vert
\eta \!-\! \eta'\vert - i \delta)^2 \; . \label{Dx^2}
\end{equation}
Note that we label the position in spacetime through the $D$-vector $x^{\mu} 
= (\eta,\vec{x})$. We see from the second term of (\ref{L}) that, for $f=0$,
the combination $a^{\frac{D-1}2} \psi$ behaves like a free, massless Dirac 
field in flat space. It follows that the propagator of $\psi$ is just a 
conformal rescaling of the flat space result,
\begin{equation}
i\Bigl[ {}_iS_j\Bigr](x;x') = (a a')^{\frac{1-D}2} \gamma^{\mu}_{ij} \, i 
\partial_{\mu} \left\{ \frac{\Gamma(\frac{D}2 -1)}{4 \pi^\frac{D}2} 
\Bigl[{\Delta x}^2(x;x') \Bigr]^{1-\frac{D}2} \right\} = 
\frac{\Gamma(\frac{D}2)}{2 \pi^\frac{D}2} \frac{(a a')^{\frac{1-D}2} (-i) 
\gamma^{\mu}_{ij} {\Delta x}_{\mu}}{[{\Delta x}^2(x;x')]^{\frac{D}2}} \; . 
\label{Sij}
\end{equation}
Here ${\Delta x}_{\mu} \equiv \eta_{\mu\nu} (x^{\nu} \!-\! x^{\prime \nu})$.
The split index notation in $i [{}_iS_j](x;x')$ indicates that the first index 
($i$) transforms according to the local Lorentz group at the first coordinate
argument ($x^{\mu}$) whereas the second index ($j$) transforms at the second 
argument ($x^{\prime \mu}$).

It has not been necessary so far to make any assumption about the form of 
the scale factor. However, the free scalar is not conformally invariant, 
which makes its plane wave mode functions depend upon $a(t)$ in a 
complicated manner. Although the solution for general $a(t)$ has recently 
been obtained \cite{TW}, the Fourier mode sum has only been worked out for 
certain cases. We therefore specialize to locally de Sitter inflation,
\begin{equation}
a(t) = e^{H t} = -\frac1{H \eta} \; .
\end{equation}
In this geometry there is a simple relation between ${\ell}(x;x')$, the 
invariant length from $x^{\mu}$ to $x^{\prime \mu}$, and the conformal
coordinate interval ${\Delta x}^2(x;x')$,
\begin{equation}
4 \sin^2\left(\frac12 H \ell(x;x') \right) = a a' H^2 {\Delta x}^2(x;x') 
\equiv y(x;x') \; .
\end{equation}
We refer to $y(x;x')$ as the de Sitter length function. The scalar 
propagator can be expressed in terms of $y \equiv y(x;x')$ and the two scale 
factors,
\begin{eqnarray}
\lefteqn{{i \Delta}(x;x') = \frac{H^{D-2}}{(4 \pi)^{\frac{D}2}} \left\{ 2^{D-4}
\Gamma(D-1) \ln(a a') - \pi {\rm cot}(\pi {\scriptstyle \frac{D}2}) 
\frac{\Gamma(D-1)}{\Gamma(\frac{D}2)} \right.} \nonumber \\
& & \hspace{3cm} \left. + \sum_{n=1}^{\infty} \Bigl[ \frac1{n} \frac{
\Gamma(D-1+n)}{\Gamma(\frac{D}2 + n)} \Bigl(\frac{y}4\Bigr)^n - \frac1{n- 
\frac{D}2} \frac{\Gamma(\frac{D}2 - 1 + n)}{\Gamma(n)} \Bigl(\frac{y}4\Bigr)^{
n - \frac{D}2}\Bigr] \right\} \; . \label{Delta}
\end{eqnarray}
Note that the homogeneous terms (i.e., $y^n$) and the constant terms have
slightly different proportionality constants from our previous expression 
\cite{OW,PTW2}. This has been done to make (\ref{Delta}) valid for regulating 
a spacetime in which $D$ will ultimately be taken to some dimension other 
than four~\footnote{We thank Ewald Puchwein for pointing out the 
improvement.}. In $D=4$, the propagator~(\ref{Delta}) reduces to
a more familiar, elementary form,
\begin{equation} 
 i \Delta(x;x') \; \stackrel{D\rightarrow 4}{\longrightarrow} \;
     \frac{H^2}{4\pi^2}\bigg\{
            	              \frac{\eta\eta'}{\Delta x^2}
                            - \frac 12 \ln(H^2\Delta x^2) 
                            -\frac 14 + \ln(2) 
        	        \bigg\}
\,.
\end{equation}

\begin{figure}[htbp]
\vskip .1in
\leftline{\hskip 2cm\epsfig{file=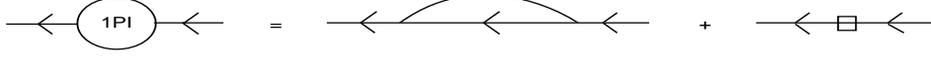, width=7.0in,height=7.0in }}
\vskip -6.5in
\caption{ {\it The one loop fermion self energy.} Fermion lines have an
arrow, scalar lines do not. The final diagram gives the contribution of 
field strength renormalization. \label{fig1} }
\end{figure}

The diagrammatic representation of the one loop fermion self energy is given 
in Fig.~\ref{fig1}. Computing it is simple: just multiply the various vertices 
and propagators,
\begin{equation}
-i \Bigl[ \hbox{}_i \Sigma_j\Bigr](x;x')\! = \!\Bigl(\! -i f a^D \delta_{ik} 
\Bigr) i\Bigl[{}_kS_{\ell}\Bigr](x;x') \Bigl(\! -i f a^{\prime D} \delta_{\ell 
j} \Bigr) i \Delta(x;x') + i {\delta Z}_2 (a a')^{\frac{D-1}2} \gamma^{\mu}_{i
j} \, i \partial_{\mu} \delta^D(x-x') \; . \label{1loop}
\end{equation}
To renormalize in four dimensions, set $D = 4 - \epsilon$ and then expand in 
$\epsilon$. Only terms which are not integrable over $d^4x'$ need to have 
$\epsilon$ arbitrary,
\begin{eqnarray}
\lefteqn{-f^2 (a a')^D i\Bigl[{}_iS_j\Bigr](x;x') i\Delta(x;x') = \frac{f^2 
(a a')^{\frac32}}{2^3 \pi^{4 - \epsilon}} \Gamma\Bigl(2- \frac{\epsilon}2
\Bigr) \Gamma\Bigl(1 - \frac{\epsilon}2\Bigr) \frac{i \gamma^{\mu}_{ij} 
{\Delta x}_{\mu}}{({\Delta x}^2)^{3 - \epsilon}} } \nonumber \\
& & \hspace{6cm} - \frac{f^2 H^2 (a a')^{\frac52}}{2^4 \pi^4} \ln\Bigl(
\frac{\sqrt{e}}4 H^2 {\Delta x}^2\Bigr) \frac{i \gamma^{\mu}_{ij} {\Delta 
x}_{\mu}}{({\Delta x}^2)^2} + O(\epsilon) \; .
\end{eqnarray}

The next step is writing the factor of ${\Delta x}_{\mu}$ as a derivative,
\begin{eqnarray}
\Gamma\Bigl(2- \frac{\epsilon}2 \Bigr) \Gamma\Bigl(1 - \frac{\epsilon}2\Bigr) 
\frac{i \gamma^{\mu}_{ij} {\Delta x}_{\mu}}{({\Delta x}^2)^{3 - \epsilon}} 
& = & -\frac14 \Gamma^2\Bigl(1 - \frac{\epsilon}2\Bigr) i \gamma^{\mu}_{ij}
\partial_{\mu} \left(\frac1{\Delta x^2} \right)^{2 - \epsilon} \; , \\
\ln\Bigl( \frac{\sqrt{e}}4 H^2 {\Delta x}^2\Bigr) \frac{i \gamma^{\mu}_{ij} 
{\Delta x}_{\mu}}{({\Delta x}^2)^2} & = & -\frac12 i \gamma^{\mu}_{ij}
\partial_{\mu} \left(\frac{ \ln\Bigl( \frac{\sqrt{e}}4 H^2 {\Delta x}^2\Bigr) 
+ 1}{\Delta x^2} \right) \; .
\end{eqnarray}
One then expresses inverse powers of ${\Delta x}^2$ as powers of $\partial^2
\equiv \eta^{\mu\nu} \partial_{\mu} \partial_{\nu}$ acting on logarithms,
\begin{equation}
\frac{ \ln\Bigl( \frac{\sqrt{e}}4 H^2 {\Delta x}^2\Bigr) + 1}{\Delta x^2}
= \frac18 \partial^2 \ln^2\Bigl( \frac{\sqrt{e}}4 H^2 {\Delta x}^2\Bigr) \; .
\end{equation}
It is at this stage that the renormalization scale $\mu$ appears and that
the ultraviolet divergence is segregated to a local term,
\begin{equation}
\left( \frac1{\Delta x^2} \right)^{2 - \epsilon} = -\frac1{32} \partial^4 
\Big[ \ln^2(\mu^2 {\Delta x}^2) - 2 \ln(\mu^2 {\Delta x^2}) \Bigr] - \frac{2 i 
\pi^{2 - \frac{\epsilon}2} \mu^{-\epsilon}}{\epsilon (1 - \epsilon) \Gamma(1 
- \frac{\epsilon}2)} \delta^D(x - x') + O(\epsilon) \; . \qquad
\end{equation}
Defining $\gamma^{\mu}_{ij} \partial_{\mu} \equiv \not{\! \partial}_{ij}$ 
and combining terms we have,
\begin{eqnarray}
\lefteqn{-f^2 (a a')^D i\Bigl[{}_iS_j\Bigr](x;x') i\Delta(x;x') = \frac{f^2 
(a a')^{\frac32}}{2^{10} \pi^4} \, i \!\! \not{\! \partial}_{ij} \partial^4 
\Bigl[ \ln^2(\mu^2 {\Delta x}^2) - 2 \ln(\mu^2 {\Delta x^2}) \Bigr] } 
\nonumber \\
& & + \frac{f^2 H^2 (a a')^{\frac52}}{2^8 \pi^4} \, i \!\! \not{\! 
\partial}_{ij} \partial^2 \ln^2\Bigl(\frac{\sqrt{e}}4 H^2 {\Delta x}^2\Bigr) + 
\frac{i f^2 \mu^{-\epsilon} (a a')^{\frac32}}{2^4 \pi^{2 - \frac{\epsilon}2}} 
\frac{\Gamma(1 - \frac{\epsilon}2)}{\epsilon (1 - \epsilon)} \, i \!\! \not{\!
\partial}_{ij} \delta^D(x - x') + O(\epsilon) \; . \qquad
\end{eqnarray}

The simplest renormalization condition is,
\begin{equation}
\delta Z_2 = -\frac{f^2 \mu^{-\epsilon}}{2^4 \pi^{2 - \frac{\epsilon}2}} 
\frac{\Gamma(1 - \frac{\epsilon}2)}{\epsilon (1 - \epsilon)} \; .
\end{equation}
We obtain the renormalized self energy by taking the limit $\epsilon 
\rightarrow 0$,
\begin{eqnarray}
\lefteqn{\Bigl[{}_i\Sigma_j^{\rm ren}\Bigr](x;x') = - \frac{f^2 (a a')^{
\frac32}}{2^{10} \pi^4} \not{\! \partial}_{ij} \partial^4 \Bigl[\ln^2(\mu^2 
{\Delta x}^2) - 2 \ln(\mu^2 {\Delta x^2}) \Bigr] } \nonumber \\
& & - \frac{f^2 (a a')^{\frac32}}{2^5 \pi^2} \ln(a a') \, i \!\! \not{\! 
\partial}_{ij} \delta^4(x - x') - \frac{f^2 H^2 (a a')^{\frac52}}{2^8 \pi^4} 
\not{\! \partial}_{ij} \partial^2 \ln^2\Bigl(\frac{\sqrt{e}}4 H^2 {\Delta x}^2
\Bigr) + O(f^4) \; . \label{IO}
\end{eqnarray}
The first term is the conformally rescaled flat space result. The second term
is the well-known contribution from the conformal anomaly. The intrinsic de 
Sitter result is the third term. It is distinguished from the other two
by its extra factor of $a a'$. 

\section{Solving the modified Dirac equation}

The self energy modifies the Dirac equation to include quantum corrections.
The modified equation takes the form,
\begin{equation}
a^{\frac32} i \! \not{\! \partial}_{ij} \Bigl(a^{\frac32} \psi_i(x) \Bigr) + 
\int d^4x' \Bigl[ \hbox{}_i \Sigma_j\Bigr](x;x') \psi_j(x') = 0 \; . \label{MD}
\end{equation}
It describes how the mode functions change due to quantum corrections. From 
these mode functions one can infer particle production, and also kinematical 
properties such as mass. 

There is some ambiguity about what it means to ``include quantum 
corrections,'' and each of the possible meanings corresponds to employing a 
different self energy functional in the modified Dirac equation (\ref{MD}). 
The in-out self energy (\ref{IO}) corresponds to a particle which is 
surrounded by free vacuum in the distant past, and for which quantum effects 
conspire to leave it surrounded by free vacuum in the distant future. This 
self energy is appropriate to flat space scattering experiments, but it is 
not very realistic for cosmological settings. A better choice for cosmology 
is to release the particle surrounded by free vacuum at some finite time and 
then let it evolve as it will. For that situation it is the Schwinger-Keldysh 
self energy which is appropriate \cite{Schwinger}.

At one loop order the Schwinger-Keldysh self energy is simple to construct
from the in-out self energy (\ref{IO}). We will give the technique without 
deriving it \cite{Jordan}. Each vertex point is endowed with a polarity $\pm$. 
In place of the conformal interval (\ref{Dx^2}) we define,
\begin{equation}
{\Delta x}^2_{++} \equiv {\Delta r}^2 - (\vert {\Delta \eta}\vert - i 
\delta)^2 \qquad , \qquad {\Delta x}^2_{+-} \equiv {\Delta r}^2 - ({\Delta 
\eta} + i \delta)^2 \; ,
\end{equation}
where ${\Delta r} \equiv \Vert \vec{x} - \vec{x}' \Vert$ and ${\Delta \eta}
\equiv \eta \!-\! \eta'$. At one loop order we do not require the $-+$ or
$--$ variations. The $++$ and $+-$ self energies are,
\begin{eqnarray}
\lefteqn{\Bigl[{}_i\Sigma_j^{\rm ren}\Bigr](x_+;x'_+) = - \frac{f^2 (a a')^{
\frac32}}{2^{10} \pi^4} \not{\! \partial}_{ij} \partial^4 \Bigl[\ln^2(\mu^2 
{\Delta x}^2_{++}) - 2 \ln(\mu^2 {\Delta x}^2_{++}) \Bigr] } \nonumber \\
& & \hspace{1cm} - \frac{f^2 (a a')^{\frac32}}{2^5 \pi^2} \ln(a a') \, i \!\! 
\not{\! \partial}_{ij} \delta^4(x \!-\! x') - \frac{f^2 H^2 (a a')^{\frac52}}{
2^8 \pi^4} \! \not{\! \partial}_{ij} \partial^2 \ln^2\Bigl(\frac{\sqrt{e}}4 H^2 
{\Delta x}^2_{++} \Bigr) \! + O(f^4) , \quad \;\; \\
\lefteqn{\Bigl[{}_i\Sigma_j^{\rm ren}\Bigr](x_+;x'_-) = + \frac{f^2 (a a')^{
\frac32}}{2^{10} \pi^4} \not{\! \partial}_{ij} \partial^4 \Bigl[\ln^2(\mu^2 
{\Delta x}^2_{+-}) - 2 \ln(\mu^2 {\Delta x}^2_{+-}) \Bigr] } \nonumber \\
& & \hspace{6.5cm} + \frac{f^2 H^2 (a a')^{\frac52}}{2^8 \pi^4} \not{\! 
\partial}_{ij} \partial^2 \ln^2\Bigl(\frac{\sqrt{e}}4 H^2 {\Delta x}^2_{+-} 
\Bigr) + O(f^4) .
\end{eqnarray}
The Schwinger-Keldysh self energy is the sum of these,
\begin{equation}
\Bigl[ \hbox{}_i \Sigma_j^{\rm SK}\Bigr](x;x') \equiv \Bigl[ \hbox{}_i 
\Sigma_j^{\rm ren}\Bigr](x_+;x'_+) + \Bigl[ \hbox{}_i \Sigma_j^{\rm ren}
\Bigr](x_+;x'_-) \; .
\end{equation}

Our explicit result for $[{}_i\Sigma_j^{\rm SK}](x;x')$ is,
\begin{eqnarray}
\lefteqn{-\frac{f^2 (a a')^{\frac32}}{2^8 \pi^3} \, i \!\! \not{\! \partial
}_{ij} \partial^4 \! \left\{\theta({\Delta \eta}) \theta({\Delta \eta} \!-\! 
{\Delta r}) \ln\Bigl[ e^{-1}\! \mu^2 ({\Delta \eta}^2 \!\!-\! {\Delta r}^2)
\Bigr] \! \right\} \!-\! \frac{f^2 (a a')^{\frac32}}{2^5 \pi^2} \ln(a a') \, 
i \!\! \not{\! \partial}_{ij} \delta^4(x \! - \! x') } \nonumber \\
& & \hspace{2cm} - \frac{f^2 H^2 (a a')^{\frac52}}{2^6 \pi^3} \, i \!\! \not{\! 
\partial}_{ij} \partial^2 \! \left\{\theta({\Delta \eta}) \theta({\Delta 
\eta} \!-\! {\Delta r}) \ln\Bigl[\frac{\sqrt{e}}4 H^2 ({\Delta \eta}^2 \!-\! 
{\Delta r}^2) \Bigr] \right\} + O(f^4) . \qquad \label{Sigma}
\end{eqnarray}
Note that it vanishes whenever the point $x^{\prime \mu}$ is outside the past
light-cone of $x^{\mu}$. This means that although the modified Dirac equation
(\ref{MD}) is nonlocal, it is nonetheless causal. Another important property
of $[{}_i\Sigma_j^{\rm SK}](x;x')$ is that each term contains only a single
gamma matrix. This means that it conserves helicity, and hence that the 
fermion remains massless.

Note also that the first term of (\ref{Sigma}), the flat space part, has the 
same number of scale factors as the tree order differential operator in 
(\ref{MD}). Perturbation theory only makes sense if the coupling constant is 
small, so we may assume $f^2 \ll 1$. This means that the flat space part of 
the one loop self energy can be ignored. So too can the conformal anomaly, 
which is only enhanced by a factor of $\ln(a)$. However, we cannot necessarily 
ignore the de Sitter correction. Although it is also down by $f^2$, it is 
enhanced by a scale factor, which becomes enormously large during inflation. 
It turns out that higher loop corrections --- which are down by more powers 
of $f$ --- can never give more than this one factor of $a$. It therefore makes 
sense to keep only the one loop de Sitter contribution of (\ref{Sigma}),
\begin{equation}
\Bigl[ \hbox{}_i \Sigma_j\Bigr](x;x') \longrightarrow - \frac{f^2 H^2 
(a a')^{\frac52}}{2^6 \pi^3} \, i \!\! \not{\! \partial}_{ij} \partial^2 
\! \left\{\theta({\Delta \eta}) \theta({\Delta \eta} \!-\! {\Delta r}) 
\ln\Bigl[\frac{\sqrt{e}}4 H^2 ({\Delta \eta}^2 \!-\! {\Delta r}^2) \Bigr] 
\right\} . 
\end{equation}
This is the ``self energy'' we will henceforth employ in the modified Dirac
equation (\ref{MD}).

Because the background is spatially translation invariant it makes sense to 
look for plane wave solutions,
\begin{equation}
\psi_i(\eta,\vec{x}) = a^{-\frac32} \chi_i(\eta,k) e^{i\vec{k} \cdot \vec{x}}
\; ,
\end{equation}
To take advantage of helicity conservation we employ chiral gamma matrices,
\begin{equation}
\gamma^0 = \left(\matrix{0 & I \cr I & 0}\right) \qquad , \qquad
\gamma^i = \left(\matrix{0 & \sigma^i \cr -\sigma^i & 0}\right) \; ,
\end{equation}
and we factor out the tree $\eta$ dependence of the 2-component helicity 
eigenstates,
\begin{equation}
\chi_i(\eta,k) = \left(\matrix{\chi_L^{\pm}(\eta,k) e^{\pm i k \eta} \cr 
\chi_R^{\pm}(\eta,k) e^{\mp i k \eta}} \right) \qquad {\rm with} \qquad 
\vec{k} \cdot \vec{\sigma} \, \chi^{\pm}_{L,R} = \pm k \, \chi_{L,R}^{\pm} \; .
\end{equation}
For the left-handed spinor the result of performing the spatial integrations
is,
\begin{equation}
\partial_0 \chi_L^{\pm}(\eta,k) + \frac{f^2 H^2}{4 \pi^2} a \int_{\eta_i
}^{\eta} \!\! d\eta' a' \chi_L^{\pm}(\eta',k) e^{\mp i 2 k {\Delta \eta}} 
\left[ \ln(H {\Delta \eta}) \!+\! \frac34 
\!+\!\frac12 \int_0^{2 k {\Delta \eta}} 
\!\!\!\! d\tau \frac{e^{\pm i \tau} \!\!-\! 1}{\tau} \right] = 0 . \label{nonL}
\end{equation}
Here $\eta_i = -1/H$ is the initial conformal time, which corresponds to
$t=0$. The analogous equation for the right-handed spinor with $\pm$ helicity
is,
\begin{equation}
\partial_0 \chi_R^{\pm}(\eta,k) + \frac{f^2 H^2}{8 \pi^2} a \int_{\eta_i
}^{\eta} \!\! d\eta' a' \chi_R^{\pm}(\eta',k) e^{\pm i 2 k {\Delta \eta}} 
\left[ 2 \ln(H {\Delta \eta}) \!+\! \frac32 \!+\! \int_0^{2 k {\Delta \eta}} 
\!\!\!\! d\tau \frac{e^{\mp i \tau} \!\!-\! 1}{\tau} \right] = 0 . \label{nonR}
\end{equation}

Recovering the full time evolution of nonlocal equations such as 
(\ref{nonL}-\ref{nonR}) requires numerical integration. However, it is 
straightforward to get the asymptotic behavior for late times. First change
the independent variable to the number of e-foldings, $N \equiv \ln(a)$. Now
redefine the dependent variable as $h_{L,R}^{\pm}(N,w)$, where $w \equiv k/H$.
The equations which result are,
\begin{eqnarray}
\lefteqn{\partial_N h_L^{\pm}(N,w) + \Bigl(\frac{f}{2 \pi}\Bigr)^2 \int_0^N 
\!\!\! dN' \, h_L^{\pm}(N',w) \exp\Bigl[\mp i 2 w (e^{-N'} \!\!\!-\! e^{-N})
\Bigr] } \nonumber \\
& & \hspace{2cm} \times \left\{\ln\Bigl[e^{-N'} \!\!\!-\! e^{-N}\Bigr] \!+\! 
\frac34 \!+\! \frac12 \int_0^{2 w (e^{-\!N'} \!\!-\, e^{-\!N})} \!\!\! d\tau 
\; \frac{e^{\pm i \tau} \!\!-\! 1}{\tau} \right\} = 0 \; . \label{newL}
\end{eqnarray}
The equations for the right-handed spinor are obtained by replacing $\mp i 2 w$
with $\pm 2 i w$ and $\pm i \tau$ with $\mp i \tau$.

To obtain the asymptotic form for large $N$ we simplify (\ref{newL}) by
dropping terms in the integrand which are exponentially small in $N$ for
fixed $N' < N$,
\begin{equation}
\partial_N h_L^{\pm}(N,w) + \Bigl(\frac{f}{2 \pi}\Bigr)^2 \int_0^N \!\!\! dN' 
\, h_L^{\pm}(N',w) e^{\mp i 2 w e^{-N'}} \left[-N' \!+\! \frac34 \!+\! 
\frac12 \int_0^{2 w e^{-\!N'}} \!\!\! d\tau \; \frac{e^{\pm i \tau} \!\!-\! 
1}{\tau} \right] \approx 0 \; . \label{simpL}
\end{equation}
At large $N$ we see that the system has the form,
\begin{equation}
\partial_N h_{\infty}(N,w) = \int_0^N \!\!\! dN' g_{\infty}(N',w) 
h_{\infty}(N',w) \; .
\end{equation}
where the function $g_{\infty}(N,w)$ is,
\begin{equation}
g_{\infty}(N,w) = \Bigl(\frac{f}{2 \pi}\Bigr)^2 \!\! \left\{N \!-\! \frac34 
\!-\! \frac12 \int_0^{2 w e^{-\!N}} \!\!\! d\tau \; \frac{e^{\pm i \tau} 
\!\!-\! 1}{\tau} \right\} e^{\mp i 2 w e^{-N}} \longrightarrow \Bigl(\frac{f}{
2 \pi} \Bigr)^2 \!\! \left\{N \!-\! \frac34 + O\Bigl(N e^{-N}\Bigr) \right\} .
\end{equation}
Because the function $g_{\infty}(N,w)$ grows, the integral is dominated by its 
upper limit and we can obtain the asymptotic form of $h_{\infty}(N,w)$ (up to 
a proportionality constant) using the WKB technique,
\begin{equation}
h_{\infty}(N,w) \longrightarrow g_{\infty}^{\!-\frac14}(N,w) \exp\Bigl[ \int^N 
\!\!\! dN' \sqrt{g_{\infty}(N',w)} \Bigr] \longrightarrow \Bigl(N - \frac34
\Bigr)^{\!-\frac14} \exp\Bigl[ \frac{f}{3 \pi} \Bigl(N - \frac34\Bigr)^{
\frac32} \Bigr] . \label{limform}
\end{equation}
Each of the various helicities and polarizations, $h^{\pm}_L(N,w)$ and 
$h^{\pm}_R(N,w)$, approaches a constant times this same limiting form
$h_{\infty}(N,w)$.

\section{The physics behind our result}

\subsection{Massless minimally coupled scalars}

Most familiar particles become (classically) conformally invariant when their
masses are taken to zero. The exceptions are gravitons and massless, minimally
coupled scalars. The Lagrangian density for the later is,
\begin{equation}
{\cal L}_{\rm MMCS} = -\frac12 \partial_{\mu} \phi \partial_{\nu} \phi 
g^{\mu \nu} \sqrt{-g} = -\frac12 a^2 \partial_{\mu} \phi \partial_{\nu}
\phi \eta^{\mu\nu} \; .
\end{equation}
The Lagrangian is the spatial integral of the Lagrangian density. Using 
Parseval's theorem to convert to Fourier space, and expressing the field in
terms of co-moving (rather than conformal) time, gives a form we can recognize,
\begin{equation}
L_{\rm MMCS} \equiv \int d^3x {\cal L}_{\rm MMCS} = \frac12 \int 
\frac{d^3p}{(2 \pi)^3} \Bigl\{ a^3(t) \vert \dot{\widetilde{\phi}}(t,\vec{p})
\vert^2 - a(t) \Vert \vec{p} \Vert^2 \vert \widetilde{\phi}(t,\vec{p}) \vert^2 
\Bigr\} .
\end{equation}
Each wave vector $\vec{p}$ represents an independent harmonic oscillator
with a time dependent mass, $m(t) \sim a^3(t)$, and frequency, $\omega(t)
= \Vert \vec{p} \Vert/a(t)$.

For the case of de Sitter inflation ($a(t) = e^{H t}$) the time dependence
of the part of the field with wave vector $\vec{k}$ takes the simple form,
\begin{equation}
\widetilde{\phi}(t,\vec{p}) = f(t,p) \alpha(\vec{p}) + f^*(t,p) 
\alpha^{\dagger}(-\vec{p}) \qquad {\rm where} \qquad f(t,p) \equiv \frac{H}{
\sqrt{2 p^3}} \Bigl(1 - \frac{i p}{a H} \Bigr) e^{i p/Ha} . \label{fmode}
\end{equation}
To understand the meaning of the canonically normalized creation and 
annihilation operators operators $\alpha(\vec{p})$ and $\alpha^{\dagger}(
\vec{p})$, note that the minimum energy in wave vector $\vec{p}$ at time $t$ 
is $\frac12 (\hbar) \omega(t)$. However, the state with this energy does not 
evolve onto itself. Indeed, this theory has no stationary states! A reasonable 
``vacuum'' is the state that was minimum energy in the distant past. This is 
known as Bunch-Davies vacuum and it is defined by $\alpha(\vec{p}) \vert 
\Omega \rangle = 0$.

For Bunch-Davies vacuum the 0-point energy in wave vector $\vec{p}$ is,
\begin{equation}
E_0(t,\vec{p}) = \frac12 a^3(t) \vert \dot{f}(t,p) \vert^2 + \frac12 a(t)
\Vert \vec{p} \Vert^2 \vert f(t,p) \vert^2 = \frac{\Vert \vec{p} \Vert}{2 a(t)}
+ \frac{H^2 a(t)}{4 \Vert \vec{p} \Vert} \; .
\end{equation}
The first term on the far right represents the irreducible, minimum energy.
The second term gives the extra energy due to inflationary particle production.
Since the energy of a particle of wave number $\vec{p}$ is $\Vert \vec{p}
\Vert/a(t)$, we can easily compute the number of particles,
\begin{equation}
N(t,\vec{p}) = \left( \frac{H a(t) }{2 \Vert \vec{p} \Vert } \right)^2 \; .
\end{equation}
As one might expect, this is much less than one at very early times, and it 
becomes of order one when the wave number just begins to satisfy the 
uncertainty bound (\ref{unc}).

\subsection{Production of fermions}

Although we have seen that inflation produces enormous numbers of massless,
minimally coupled scalars, the conformal invariance of free Dirac theory
implies that there can be no comparable, direct production of fermi\-ons. 
However, it is still possible to make fermions during inflation by allowing
them to interact with a massless, minimally coupled scalar. To see this we 
begin by giving the free field expansions of the interaction picture fields,
\begin{eqnarray}
\phi_I(t,\vec{x}) & = & \int \frac{d^3p}{(2\pi)^3} \left\{ \alpha(\vec{p}) 
f(t,p) e^{i \vec{p} \cdot \vec{x}} + \alpha^{\dagger}(\vec{p}) f^*(t,p) 
e^{-i \vec{p} \cdot \vec{x}} \right\} , \\
\psi_I(t,\vec{x}) & = & \int \frac{d^3q}{(2\pi)^3} \sum_r \left\{ 
\beta(\vec{q},r) \frac{u(\vec{q},r)}{\sqrt{2 q}} e^{iq/Ha + i \vec{q} \cdot 
\vec{x}} + \gamma^{\dagger}(\vec{q},r) \frac{v(\vec{q},r)}{\sqrt{2 q}} 
e^{-iq/Ha - i \vec{q} \cdot \vec{x}} \right\} , \\
\overline{\psi}_I(t,\vec{x}) & = & \int \frac{d^3k}{(2\pi)^3} \sum_s \left\{ 
\gamma(\vec{k},s) \frac{\overline{v}(\vec{k},s)}{\sqrt{2 k}} e^{ik/Ha + i 
\vec{k} \cdot \vec{x}} + \beta^{\dagger}(\vec{k},s) \frac{\overline{u}(
\vec{k},s)}{\sqrt{2 k}} e^{-ik/Ha - i \vec{k} \cdot \vec{x}} \right\} .
\end{eqnarray}
Note that we have absorbed a factor of $a^{\frac32}$ into $\psi_I$ and 
$\overline{\psi}_I$ so that they agree exactly with the flat space expansions
in conformal coordinates. The spinor wave functions obey the usual spin sum
formulae of massless fermions in flat space,
\begin{equation}
\sum_s u(\vec{k},s) \overline{u}(\vec{k},s) = \not{\! k} = \sum_s v(\vec{k},s) 
\overline{v}(\vec{k},s) \; .
\end{equation}
The nonzero relations of the operator algebra are,
\begin{eqnarray}
\Bigl[\alpha(\vec{p}),\alpha^{\dagger}(\vec{p}^{~\prime})\Bigr] & = & (2\pi)^3 
\delta^3( \vec{p} - \vec{p}^{~\prime}) \; , \\
\Bigl\{\beta(\vec{k},s),\beta^{\dagger}(\vec{k}',s')\Bigr\} & = & (2\pi)^3 
\delta^3(\vec{k} - \vec{k}') \delta_{ss'} = \Bigl\{\gamma(\vec{k},s),\gamma^{
\dagger}(\vec{k}',s')\Bigr\} \; .
\end{eqnarray}

We are studying the same Yukawa theory as in sections II and III and, as 
before, we release the universe in free Bunch-Davies vacuum at $t=0$,
\begin{equation}
\alpha(\vec{p}) \vert \Omega \rangle = 0 = \beta(\vec{k},s) \vert \Omega 
\rangle = \gamma(\vec{q},r) \vert \Omega \rangle \; .
\end{equation}
Therefore the operator which maps the various free fields into interacting 
fields is,
\begin{equation}
U(t) = T\left\{ \exp\Bigl[-i f \!\! \int_0^t \!\! dt' \!\! \int \!\! d^3x 
\phi_I(t',\vec{x}) \overline{\psi}_I(t',\vec{x}) \psi_I(t',\vec{x}) \Bigr] 
\right\} \; . \label{U}
\end{equation}
Although Heisenberg states such as $\vert \Omega \rangle$ do not change with
time, what these states mean is determined by operators, which do change with
time. In particular, the mapping,
\begin{equation}
a^{\frac32}(t) \psi(t,\vec{x}) = U^{\dagger}(t) \psi_I(t,\vec{x}) U(t) \; ,
\end{equation}
suggests a reasonable definition for the operator which annihilates fermions 
at time $t$,
\begin{equation}
\beta(t,\vec{k},s) \equiv U^{\dagger}(t) \beta(\vec{k},s) U(t) \; .
\end{equation}
We therefore arrive at an expression for the probability that $\vert \Omega 
\rangle$ contains a fermion of wave number $\vec{k}$ and spin $s$ at time $t$,
\begin{equation}
P_{\beta}(t,\vec{k},s) (2\pi)^3 \delta^3(\vec{k} - \vec{q}) \delta_{rs} =
\Bigl\langle \Omega \Bigl\vert \beta^{\dagger}(t,\vec{q},r) \beta(t,\vec{k},s)
\Bigr\vert \Omega \Bigr\rangle = \Bigl\langle \Omega \Bigl\vert U^{\dagger}(t)
\beta^{\dagger}(\vec{q},r) \beta(\vec{k},s) U(t) \Bigr\vert \Omega \Bigr\rangle
\; . \label{prob}
\end{equation} 

Expression (\ref{prob}) is straightforward to evaluate to lowest order. One
first commutes the creation and annihilation operators through the factors of
$U^{\dagger}(t)$ and $U(t)$,
\begin{equation}
P_{\beta}(t,\vec{k},s) (2\pi)^3 \delta^3(\vec{k} - \vec{q}) \delta_{rs} =
\Bigl\langle \Omega \Bigl\vert \Bigl[\beta(\vec{q},r),U(t)\Bigr]^{\dagger}
\Bigl[\beta(\vec{k},s),U(t)\Bigr] \Bigr\vert \Omega \Bigr\rangle \; . 
\end{equation}
From the definition (\ref{U}) of $U(t)$, and the free field expansions,
we find,
\begin{equation}
\Bigl[\beta(\vec{k},s),U(t)\Bigr] = \frac{-i f}{\sqrt{2k}} \int_0^t \!\! dt'
\!\! \int \!\! d^3x \phi_I(t',\vec{x}) \overline{u}(\vec{k},s) \psi(t',\vec{x})
e^{-i k/H a' - i \vec{k} \cdot \vec{x}} + O(f^2) \; .
\end{equation}
The free field expansions imply,
\begin{eqnarray}
\lefteqn{\Bigl\langle \Omega \Bigl\vert \phi_I(t'',\vec{y}) \overline{\psi
}_I(t'',\vec{y}) u(\vec{q},r) \phi_I(t',\vec{x}) \overline{u}(\vec{k},s) 
\psi_I(t',\vec{x}) \Bigr\vert \Omega \Bigr\rangle = \int \!\! \frac{d^3p}{(2
\pi)^3} f^*(t',p) f(t'',p) e^{-i \vec{p} \cdot (\vec{x} - \vec{y})} } 
\nonumber \\
& & \hspace{3cm} \times \int \!\! \frac{d^3\ell}{(2\pi)^3} \frac1{2 \ell} 
\sum_{s'} \overline{v}(\vec{\ell},{s'}) u(\vec{q},r) \overline{u}(\vec{k},s) 
v(\vec{\ell},s') e^{-i \ell/H a' + i \ell/H a'' - i \vec{\ell} \cdot (\vec{x}
- \vec{y})} \; , 
\end{eqnarray}
The integrals over $\vec{x}$ and $\vec{y}$ enforce $\vec{q} = \vec{k}$ and
$\vec{\ell} = -(\vec{k} + \vec{p})$. With these identifications we can set 
$r=s$ and spin average to eliminate the fermion wave functions,
\begin{equation}
\frac12 \sum_{s,s'} \overline{v}(\vec{\ell},{s'}) u(\vec{k},s) \overline{u}(
\vec{k},s) v(\vec{\ell},s') = \frac12 {\rm Tr}\Bigl[\not {\! \ell} \not {\! k}
\Bigr] = 2 \Bigl[ \Vert \vec{k} + \vec{p} \Vert k + (\vec{k} + \vec{p}) \cdot 
\vec{k} \Bigr] \; .
\end{equation}
Putting everything together gives,
\begin{equation}
P_{\beta}(t,\vec{k},s) = \frac{f^2}{2k} \!\! \int_0^t \!\! dt' \!\! \int_0^t
\!\! dt'' \!\! \int \!\! \frac{d^3p}{(2\pi)^3} f^*(t',p) f(t'',p) e^{-i (
\frac1{Ha'} - \frac1{Ha''}) (\Vert \vec{p} + \vec{k} \Vert + k)} \Bigl[k \!+\!
\frac{(\vec{k} + \vec{p}) \cdot \vec{k}}{\Vert \vec{k} \!+\! \vec{p} 
\Vert}\Bigr] + O(f^4) . \label{final}
\end{equation}

The time dependence of the integrand of (\ref{final}),
\begin{equation}
f^*(t',p) f(t'',p) e^{-i (\frac1{Ha'} - \frac1{Ha''}) (\Vert \vec{p} + \vec{k} 
\Vert + k)} = \frac{H^2}{2 p^3} \Bigl(1 + \frac{i p}{Ha'}\Bigr) \Bigl(1 - 
\frac{ip}{H a''}\Bigr) e^{-i (\frac1{Ha'} - \frac1{Ha''}) (\Vert \vec{p} + 
\vec{k} \Vert + p + k)} \; , \label{integrand}
\end{equation}
indicates that the lowest order effect is driven by the physical process in
which a scalar and a fermion-anti-fermion pair emerge from the vacuum. In
flat space quantum field theory this process would be forbidden by energy 
conservation. It is crucial to note that this is not put in by hand: even
in flat space one gets an expression analogous to (\ref{final}), but it goes
to zero as the time $t$ tends to infinity. What drives the flat space 
expression to zero is the fact that no value of the scalar wave vector
$\vec{p}$ gives zero for the total flat space energy, $\Vert \vec{p} + \vec{k} 
\Vert + p + k$. Hence the temporal integrations result in rapid oscillations 
that tend to cancel. As $t$ grows, the result approaches an energy delta 
function that cannot be saturated. In our locally de Sitter background the 
inverse scale factors redshift the physical energy to zero so rapidly that 
there are only a finite number of oscillations.

Note the essential role played by the breaking of conformal invariance.
Since free fermions are conformally invariant, and the Yukawa interaction
is not inconsistent with conformal invariance, the effect of conformal 
breaking is concentrated in the scalar wave function, $f(t,p)$. Had we instead 
used a conformally coupled scalar the integrand (\ref{integrand}) would go to, 
\begin{equation}
f^*(t',p) f(t'',p) e^{-i (\frac1{Ha'} - \frac1{Ha''}) (\Vert \vec{p} + \vec{k} 
\Vert + k)} \longrightarrow \frac1{2 a' a'' p} e^{-i (\frac1{Ha'} - 
\frac1{Ha''}) (\Vert \vec{p} + \vec{k} \Vert + p + k)} \; , \label{newgrand}
\end{equation}
In this case the inverse scale factors suppress the result.

To give a good approximate computation of $P_{\beta}(t,\vec{k},s)$ we note,
from the preceding discussion, that scalar wave numbers with $p \gtwid H A
\equiv H {\rm min}(a',a'')$ engender oscillations in the energy exponent. We 
can therefore cut the scalar wave number off above this limit. (The infrared 
cutoff of $p \ge H$ derives from the finite coordinate range of our spatial 
manifold $T^3$ \cite{OW}.) The same energy exponent also gives rise to 
oscillations when $k \gtwid H a'$ or $k \gtwid H a''$, which cuts off the 
lower limits of the temporal integrations at $t',t'' \ge T = \ln(w)/H$, where
$w \equiv k/H$. With these cutoffs in place we can neglect terms that fall 
off like $1/a'$ or $1/a''$,
\begin{equation}
P_{\beta}(t,\vec{k},s) \longrightarrow \frac{f^2}{k} \!\! \int_T^t \!\! dt'
\!\! \int_T^t \!\! dt'' \!\! \int_H^{H A} \!\!\!\! dp \, p^2 \frac{H^2}{2 p^3}
\int \!\! \frac{d^2\widehat{p}}{(2\pi)^3} \left[k + \frac{(\vec{k} + \vec{p})
\cdot \vec{k}}{\Vert \vec{k} + \vec{p} \Vert} \right] + O(f^4) \; .
\end{equation}
The angular integration is straightforward,
\begin{equation}
\int \!\! \frac{d^2\widehat{p}}{(2\pi)^3} \left[k + \frac{(\vec{k} + \vec{p})
\cdot \vec{k}}{\Vert \vec{k} + \vec{p} \Vert} \right] = \frac{k}{2\pi^2}
\left\{ \theta(k-p) \Bigl[2 - \frac{p^2}{3 k^2}\Bigr] + \theta(p-k) \Bigl[1
+ \frac{2 k}{3 p}\Bigr] \right\} .
\end{equation}
Now change variables from time to e-foldings, and take advantage of the 
symmetry between $t'$ and $t''$,
\begin{equation}
P_{\beta}(t,\vec{k},s) \longrightarrow \frac{f^2}{2 \pi^2} \!\! \int^N_{\ln w}
\!\!\!\!\!\! dN' \!\! \int^{N'}_{\ln w} \!\!\!\!\!\! dN'' \!\! \int_H^{H 
e^{N''}} \!\! \frac{dp}{p} \left\{ \theta(k\!-\!p) \Bigl[2 \!-\! \frac{p^2}{3 
k^2}\Bigr] \!+\! \theta(p\!-\!k) \Bigl[1 \!+\! \frac{2 k}{3 p}\Bigr] \! 
\right\} + O(f^4) .
\end{equation}
Performing the integrations and retaining only the leading terms for $N \gg 
\ln(w) \gg 1$, we obtain,
\begin{eqnarray}
P_{\beta}(t,\vec{k},s) & \longrightarrow & \frac{f^2}{2 \pi^2} \!\! \int^N_{
\ln w} \!\!\!\!\!\! dN' \!\! \int^{N'}_{\ln w} \!\!\!\!\!\! dN'' \Bigl\{N''
+ \ln(w) + \frac12 + \dots \Bigr\} + O(f^4) \; , \\
& \longrightarrow & \frac{f^2}{12 \pi^2} \Bigl\{ N^3 + \dots \Bigr\} + O(f^4)
\; .
\end{eqnarray}
As one might have expected, our leading order result for 
$P_{\beta}(t,\vec{k},s)$ is independent of $s$ and $w$. 
Upon taking a square root of $P_{\beta}(t,\vec{k},s)$,
one gets a quite good approximation for the exponent of the WKB mode
function~(\ref{limform}) we found in Section III.

\section{Discussion}

We have found that the effect of inflation on Yukawa-coupled, massless 
fermions is in some ways opposite to the effect on the photons of massless
scalar QED. Whereas the photon becomes massive, but suffers no significant
particle production \cite{PW2}, we find that fermions remain massless but
experience enormous particle production. The physical mechanism is simple:
inflation alters the constraint of energy conservation so that three massless
particles can spontaneously emerge from the vacuum. The Yukawa interaction
mediates this for a scalar and a fermion-anti-fermion pair, and the process
has an appreciable amplitude when the scalar is minimally coupled.

The mathematical analysis for Yukawa-coupled fermions is quite different 
from that required for photons in scalar QED. Although the mode equations 
of both systems are nonlocal, the integrals of the fermion system are 
dominated by their upper limits because the fermion mode function grows.
This means that we need only drop exponentially suppressed terms and then
differentiate to extract a local equation for the asymptotic form. The photon 
mode equations were much more subtle because the photon mode function 
approaches a nonzero constant --- as do classical photons --- and the 
asymptotic effect of vacuum polarization shows up in the increasingly rapid
oscillations around this value.

Of course fermions obey the Pauli exclusion principle, so at most one particle
can be put in each momentum and spin state. The amplitude of the wave function
cannot really grow past that point. Therefore the rate of production for any
particular momentum and spin state must go to zero as the probability of this
state being occupied approaches unity. This phenomenon is known as {\it
Pauli blocking}. We do not see Pauli blocking because we are using only the
one loop self energy. We expect Pauli blocking to appear at two loop order.

\section*{Acknowledgements}

We are grateful to H. J. Monkhorst for posing the question which led to
this study, and to Bj\"{o}rn Garbrecht and Shun-Pei Miao for useful conversations.
This work was partially supported by NSF grant
PHY-244714 and by the Institute for Fundamental Theory at the University of
Florida.

\section*{Addendum: Erratum for ``Production of Massless Fermions during Inflation''
         }

\begin{center}
\centerline{\bf Abstract}
We correct a sign error in the effective Dirac equation solved 
in~Ref.\cite{ProkopecWoodard:2003}. We discuss the changes this implies and
their physical significance.
\end{center}

In Ref.~\cite{ProkopecWoodard:2003} we calculated the one-loop self-energy
for massless fermions which couple {\it via} a Yukawa interaction term
to a massless, minimally coupled scalar during de Sitter inflation. 
However, within the system of conventions we used, the self-energy should 
contribute oppositely to the local derivative term in the quantum-corrected
Dirac equation. The correct equation is,
\begin{equation}
  a^{3/2}i\partial\!\!\!\slash (a^{3/2} \psi(x)) 
     - \int d^4x^{\,\prime} \Sigma_{\rm ret}(x,x^\prime) \psi(x^\prime) 
     = 0
\,.
\label{modified:Dirac}
\end{equation}
Here $a=-1/(H\eta)$ denotes the scale factor during de Sitter inflation,
$\Sigma_{\rm ret}$ is the retarded self-energy of the Schwinger-Keldysh
formalism, and $\psi$ is the fermionic wave function. This is how equation~(29) 
in~\cite{ProkopecWoodard:2003} should have read.

The correct sign makes a simple and profound change in our analysis. The
conformally rescaled, chiral components of spatial plane wave mode functions
obey the equation,
\begin{equation}
 \partial_N h_L^{\pm}(N,w) 
  + \Bigl(\frac{f}{2 \pi}\Bigr)^2 \int_0^N \!\!\! dN' 
    \, h_L^{\pm}(N',w) e^{\mp i 2 w e^{-N'}} \left[N' \!-\! \frac34 \!-\! 
    \frac12 \int_0^{2 w e^{-\!N'}} \!\!\! d\tau \; \frac{e^{\pm i \tau} \!\!-\! 1}{\tau} \right]
  \approx 0 
\, .
\label{simpL:B}
\end{equation}
where $w = k/H$, $N= \ln(a)$, $f$ denotes the Yukawa coupling, and the 
approximation is that some exponentially small terms have been neglected.
This should replace equation~(38) in~\cite{ProkopecWoodard:2003}.
In the late time limit of large $N = \ln(a)$, both helicity components 
approach the same declining oscillatory form,
\begin{equation}
 h^{\pm}_{L,R} \;\stackrel{N\rightarrow\infty}{\longrightarrow} \;
    \Big(N-\frac{3}{4}\Big)^{-1/4}
   \exp\bigg(\pm {\rm i} \frac{f}{3\pi}\Big(N-\frac{3}{4}\Big)^{3/2}\bigg)
\,.
\label{h Airy:asymptotic}
\end{equation}
This replaces equation~(41) in~\cite{ProkopecWoodard:2003}.

The correct asymptotic solution (\ref{h Airy:asymptotic}) stands in sharp
contrast to the faster-than-exponential growth which pertains with the sign
error! The physical interpretation of the corrected result seems to be that
the inflationary production of massless, minimally coupled scalars drives
the scalar field strength away from zero, which affects super-horizon fermions
like a mass term. For a more complete study of the mass generation mechanism 
in the case when the scalar field is light and nearly minimally coupled we 
refer to Ref.~\cite{GarbrechtProkopec:2006}.

Declining oscillatory behavior very similar to (\ref{h Airy:asymptotic})
is observed when studying late time dynamics of super-Hubble 
photons~\cite{PW2}. So the sign correction brings our results 
for Yukawa into pleasant conformity with what charged scalars do to photons 
during inflation~\cite{PTW1,PW1,PTW2,ProkopecPuchwein:2003}.
In each case inflationary particle production forces the scalar away from
zero, whereupon the effect on other fields is qualitatively similar (but not
in every detail!) to what happens for nonzero scalar backgrounds in 
symmetry breaking.

Note that the sign change reported here does not affect either the one loop
Yukawa scalar self-mass-squared or its impact on plane wave scalar mode 
functions~\cite{Duffy:2005ue}. So the conclusion of that study is still
that the scalar cannot develop a large enough mass rapidly enough to
affect what it does to the fermion. Nor does the sign change reported here
affect the recent computation of quantum gravitational corrections to the 
one loop self-energy of massless fermions in Dirac + 
Einstein~\cite{Miao:2005am}. Of course our sign change very much {\it does} 
affect the impact these corrections have on fermion mode functions~\cite{MiaoWoodard:2006}!

\end{document}